# Seizure prediction with long-term iEEG recordings: What can we learn from data nonstationarity?

Hongliu Yang*, Matthias Eberlein, Jens Müller, and Ronald Tetzlaff
TU Dresden, Faculty of Electrical and Computer Engineering, Institute of Circuits and Systems,
01062 Dresden, Germany
* Email: hongliu.yang@tu-dresden.de

*Abstract*—Repeated epileptic seizures impair around 65 million people worldwide and a successful prediction of seizures could significantly help patients suffering from refractory epilepsy. For two dogs with yearlong intracranial electroencephalography (iEEG) recordings, we studied the influence of time series nonstationarity on the performance of seizure prediction using in-house developed machine learning algorithms. We observed a long-term evolution on the scale of weeks or months in iEEG time series that may be represented as switching between certain meta-states. To better predict impending seizures, retraining of prediction algorithms is therefore necessary and the retraining schedule should be adjusted to the change in meta-states. There is evidence that the nature of seizure-free interictal clips also changes with the transition between meta-states, which has been shown relevant for seizure prediction.

## I. Introduction

The permanent uncertainty is one of the biggest burdens for people with epilepsy [1]. Driven by the advances in data analytics with machine learning, a remarkable progress in the prediction of impending seizure events has been made in the last years [2]. The evaluation of new methods, however, is treacherous because it is mostly conducted on short-term recordings (here *short-term* means data of presurgical diagnostics with a duration up to two weeks) and hence the outcome was over-optimistic [3]–[7].

Much more meaningful is an evaluation on *long-term* data such as those recently recorded from the NeuroVista clinical trial [8] and and made available to the public community [9]–[11]. But as a result, we observe that the performance of the algorithms is far from optimal and even ensembling of different methods does not lead to an improvement [12].

In order to understand the reasons for current limitations, we evaluate a selected algorithm on long-term *continuous* intracranial electroencephalography (iEEG) recordings and demonstrate how non-stationarities in the data influence the seizure-prediction performance. To do so, we investigate the effect of a systematic retraining of the model's parameters on its performance. In order to support the future design of predictive systems, all investigations are performed in a pseudo-prospective manner.

Table I
CHARACTERISTICS OF DATASET. EACH DATASET COMPRISES 16 CHANNELS WITH A SAMPLING RATE OF 400 HZ.

|       | seizures (lead.) | recording span | recorded data |
|-------|------------------|----------------|---------------|
| dog A | 45(40)           | 476 d          | 344 d         |
| dog B | 84(18)           | 452 d          | 214 d         |

Table II
SPLITS OF DATA FOR ALGORITHM DEVELOPMENT AND EVALUATION. EACH TIME PERIOD, I.E. THE RECORDING BETWEEN TWO ADJACENT SPLIT POINTS, CONTAINS A SINGLE SEIZURE CLUSTER (SEE FIG. 1). FOR THE $i$-TH TRAIN SETUP ALL DATA UNTIL THE $i$-TH SPLIT POINT WERE USED FOR TRAINING.

|       | split times / d |    |     |     |     |     |     |
|-------|-----------------|----|-----|-----|-----|-----|-----|
| index | 0               | 1  | 2   | 3   | 4   | 5   | 6   |
| dog A | 24              | 50 | 275 | 333 | 362 | 404 | 434 |
| dog B | 41              | 72 | 95  | 350 | 408 | 426 | 436 |

## II. Method

### A. Data set

We used iEEG recordings of two dogs (I004_A0002 and I004_A0003) shared by the collaborative initiative ieeg.org. Information on the data set is summarized in Tab. I. For further details, refer to [9], [13]. The data set was recorded by an implanted NeuroVista seizure advisory system. Irrespective the recording gaps its time span is *more than one order of magnitude longer* than the two-week contemporary standards.

For both dogs seizures were found to occur in clusters, i.e. the time separation between subsequent events in the same cluster is less than 12 hours while the inter-cluster separation is on the scale of weeks or months. To investigate time series non-stationarity, the complete recording was split into time periods with each containing a single seizure cluster. The splitting points are listed in Tab. II. Additionally, the timeline layout of data characteristics including seizures, data gaps, and splits is presented in Fig. 1.

### B. Detection of seizure precursors

A fundamental hypothesis of seizure prediction is that seizures come with precursors, i.e. the preictal cortical dynamics shortly before a seizure onset is different from interictal states far away from any event [14], [15]. The detection of seizure precursors is then performed by classifying iEEG segments into preictal/interictal classes.. Here we used the definition that the one-hour interval up to 5 min before an onset is preictal and interictal is one week away from any seizure [2], [9]. Continuous recordings are segmented into 10 min clips for further processing.

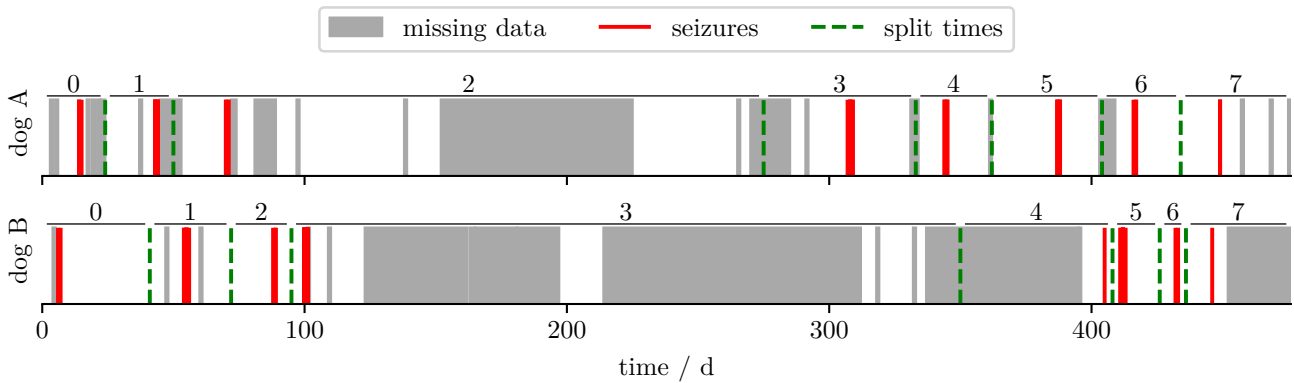

Figure 1. Temporal visualization of available data, seizures and splits. The different time periods used for training and testing are indexed as well. For the coordinate of splits, refer to Tab. II.

For an automated classification of iEEG clips, we apply multilayer perceptrons (MLP) with hand-crafted features as input: 1) Band power spectrum has shown its success in recent Kaggle competitions [2], [9]. 2) Autoregressive models were used to extract sequential information in EEG signals [16]. 3) Mean phase coherence is able to probe the weak nonlinear correlation among channels [17]. All three features are used jointly in this study and are expected to provide complementary information on seizure dynamics.

All preictal and interictal clips until a predetermined split point were used to train a MLP model while the time periods beyond this point were used for testing. Considering the strong imbalance between preictal and interictal classes, we used the precision-recall (PR) area-under-curve value to evaluate precursor detection performance in addition to the usual receiver operating characteristic (ROC) AUC value. Our MLP model has three hidden layers with 16, 8 and 4 neurons, and each is followed by a batch normalization layer, respectively. ReLU is used for all layers except a sigmoid function for the last layer. For a reliable statistical evaluation we trained 100 MLP models with different initial weights and reported the mean AUC values.

The choice of a feature-based method with MLP is based on our recent observation that such a simple method could achieve the similar state-of-the-art performance as more sophistical deep-learning counterparts [18]–[20]. Moreover, both methods deliver false predictions in coherent way [19], which is in consistence with the finding that the ensembling of top algorithms can not improve the performance further [12]. For our treatment of super-long EEG recordings in this study we use therefore the simple feature-based method with MLP to save the computational resource, and it is expected the similar behavior will be observed for other models including the deep networks in response to the long-term nonstationarity studied here.

### C. Setting seizure alarms

With the preictal probability of each data clip provided by the MLP model, a moving-window average with size $t_w$ was applied to suppress jitters. An alarm with duration $t_d$ was raised after an intervention delay $t_i$ if the smoothed preictal probability is beyond a threshold $p_{th}$. Here we fixed $t_i = 300$ s and optimized other parameters $t_d$, $p_{th}$, $t_w$ via a genetic algorithm [21] using the time period right after the train set for the MLP model. The performance of an alarm setting was evaluated by the time in warning (TIW) at 100 % sensitivity. Here TIW is the ratio of total time in alarm and the duration of used recording.

### III. Results and discussion

#### A. Performance of precursor detection and grouping of seizure clusters

The performance of preictal/interictal classification for dog A is shown in Fig. 2. To mimic the scenario of a real-time clinical trial we shifted gradually the train/test split point to include more and more data in training. An obvious finding is that test scores of a given time period could change significantly with varying the train set, which implies the nonstationarity of iEEG recordings considered. Moreover, for most cases better scores were obtained for a time period when it came closer to the train set. Therefore, to better predict the impending seizures a retraining with very recent iEEG recordings is necessary.

Another insight from Fig. 2 is the grouping of seizure clusters and the associated time periods. For dog A the eight seizure clusters fall roughly into four groups (0,1) (2) (3-6) (7). When some seizures of a group were used for training, higher AUC scores were obtained for the time periods containing remaining seizures in the group, i.e.

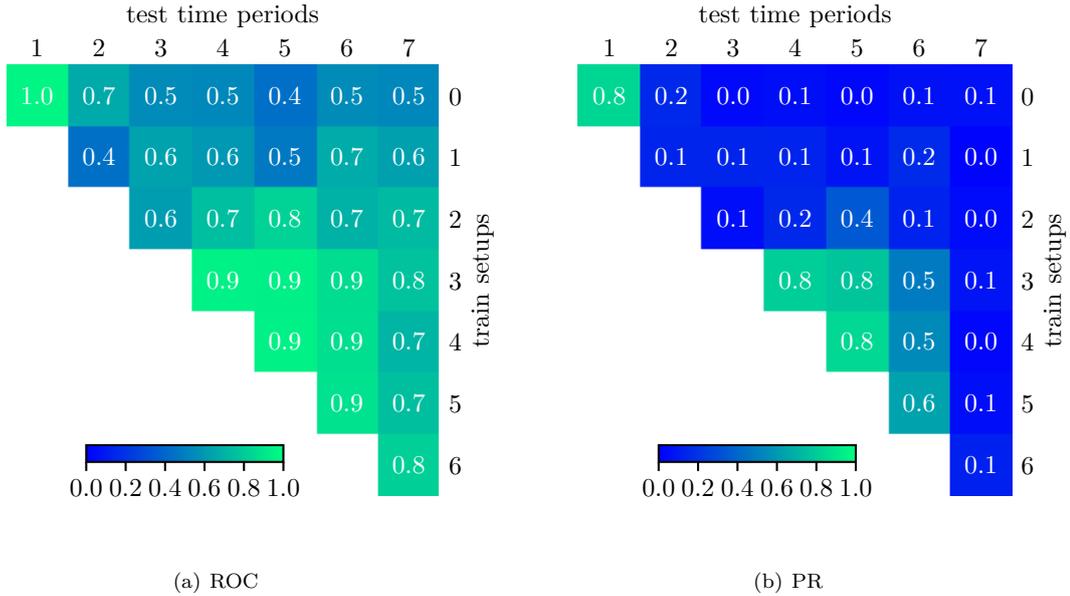

(a) ROC  (b) PR

Figure 2. Performance of classifying selected iEEG clips into interictal/preictal classes as characterized by (a) ROC and (b) PR AUC values for dog A for different combinations of train- and test periods. The indices for the respective sets are introduced in Tab. II and Fig. 1. Here a larger AUC value (green color) indicates a better performance.

seizure clusters in a group and the corresponding time periods are highly similar in nature. For example, when the algorithm is trained with all data up to split point 3 (train setup 3), the test scores of periods 4-6 are remarkably high. In contrast, for groups unseen in training the test performance is worse, see for instance the group (3-6) for train setups 0-2. This evidences the difference between seizure groups. The grouping behavior indicates the existence of meta-states for long-term cortical dynamics and the time evolution can be represented as the switching between these states.

A comparison of ROC and PR scores in Fig. 2 found that they show a similar trend of variation but the contrast of performance difference is emphasized more in PR scores. In the following, we use the PR score only.

### B. Change in interictal states and the relevance to seizure prediction

It is not expected that the nonstationarity of cortical dynamics is only reflected in the change of seizures or their precursors although that has always been the focus in seizure prediction in the past. To detect the possible change in interictal states and to demonstrate its effect on seizure prediction we successively changed the train set for dog A by including more and more interictal samples to see how the performance of precursor detection varies. As shown in Fig. 3 (a-c), with shifting the split point from day 85, to day 155, and then to day 175 the predicted preictal probability changes dramatically. With the noisy variation profile in (a) clear spike structures emerge when going from (b) to (c). Here the performance improvement for time periods 3-6 is merely due to the inclusion of interictal states recorded between day 85 and day 275.

To exclude the possibility that the performance change is caused by the increase of the amount of interictal states in train set only, we designed an alternative train setup as shown in Fig. 3 (d). Compared to (b) it has the same preictal states and a similar amount of interictal states in the train set. The outcome of precursor detection with the setup has, however, no detectable improvement over (a). Corresponding to such visual changes the PR AUC scores of (a-d) are 0.06, 0.10, 0.20 and 0.03, respectively. The results in (a-d) showed clearly that interictal states between day 85 and day 275 contain information relevant for the detection of precursors of seizure clusters 3-6. It demonstrates the dynamical change in the nature of interictal states and its importance for seizure prediction.

As a partial support to the discovered change in interictal states we show in Fig. 3 (e) an example component of our features after processed by principle component analysis (PCA) and independent component analysis (ICA) methods. With PCA we keep only first 10 principal components which amounts to 96 % variance of the original signal. One can see clearly: 1) the difference between time periods 0-1, 3-6, and 7; 2) the drift in time period 2 and its approaching to stationary state after the gap beginning at day 155; and 3) feature values in the time period 2 after day 155 being close to those in time periods 3-6. This explains on one hand the grouping of seizure clusters and on the other hand the change in nature of interictal states and why interictal states between day 85 and day 275 can help to detect precursors of seizure clusters 3-6.

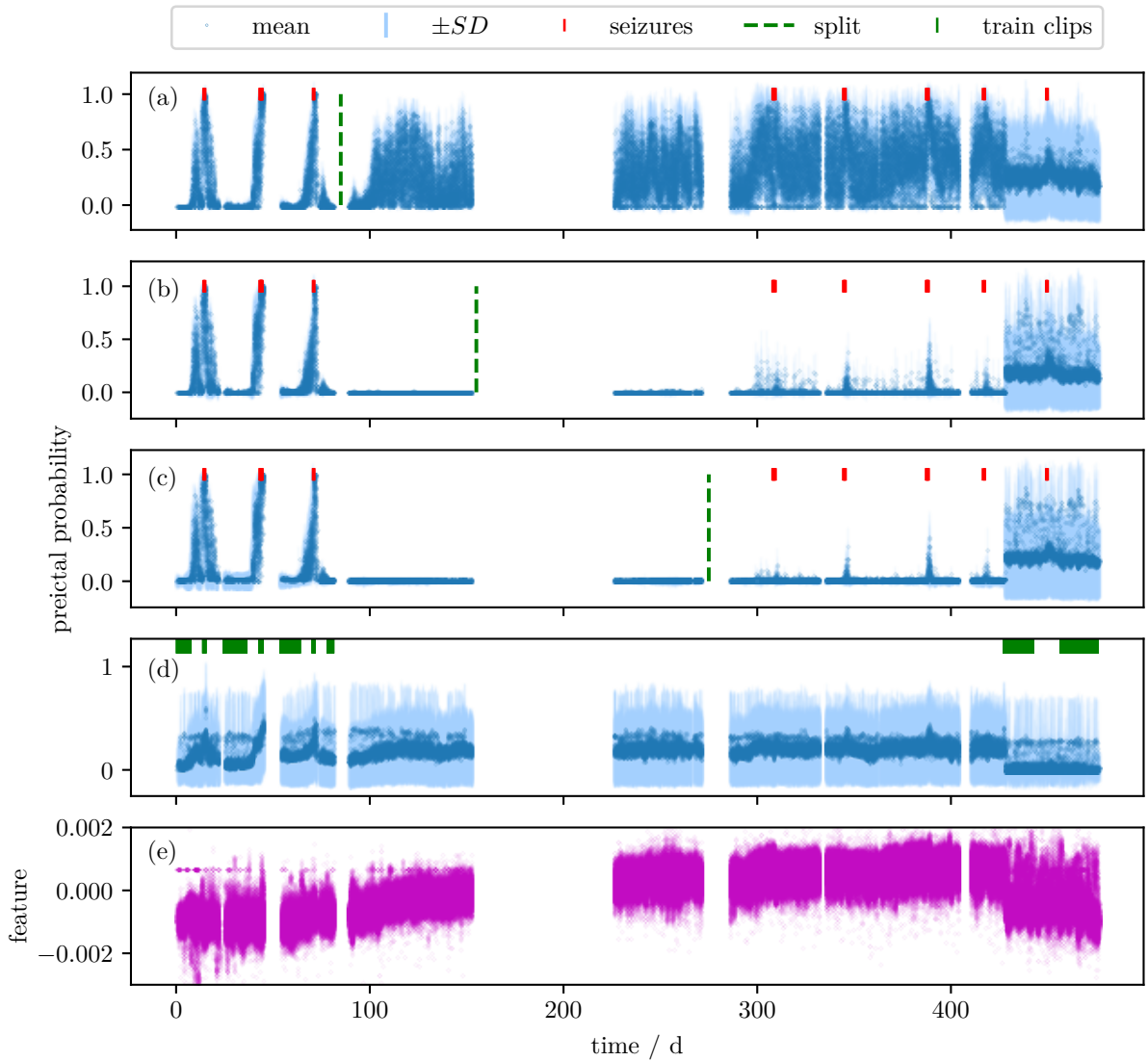

Figure 3. (a-d) Probability of being preictal clips as classified by MLP model, and (e) one component of features processed with PCA and ICA methods. Split of train/test sets is marked by a green dash line in (a-c) while the train clips are marked by the green strips in (d). Corresponding to the visual change in (a-d) the PR AUC scores are 0.06, 0.10, 0.20 and 0.03, respectively.

### C. Variation of precursor detection performance for dog B

The reported results are not specific to dog A. As shown in Fig. 4 for dog B the performance of precursor detection normally is the highest when the test time period is adjacent to the train set. There are exceptions, that could be explained by the grouping of seizure clusters and the corresponding time periods. For dog B one can identify a group structure (0-2, 5-7) (3,4). Note that two periods belonging to the first group can be separated by months, which implies the recurrent nature of the meta-state dynamics of this subject. With train setups 3 and 4 the inclusion of states from the group (3,4) leads to the lowering of performance scores for time periods 5-7 as compared to train setups 1 and 2. This has an important implication for the retraining of precursor detection algorithms: It is not the amount of train data which helps but rather the nature of the data. To be precise, trained with data (only) from the same group/meta-state as the test set will be the optimal choice.

### D. Variation of performance of seizure alarms

To characterize the performance of seizure alarms raised we present the TIW values at 100 % sensitivity for dog A in Fig. 5. Validation scores for the time period used to optimize the parameters of alarm setting are at the left side of the green line. The variation of performance can be explained by the reported grouping of seizure clusters and the corresponding time periods. Better scores are obtained when the test time period is in the same group as the validation time period used for parameter optimization, see for instance scores of time periods 4-6 for train setups 2-4. Otherwise, the performance is far worse, for example the time period 2(3) for the train setup 0(1). The high scores for time period 7 are unexpected and this deserves a close investigation.

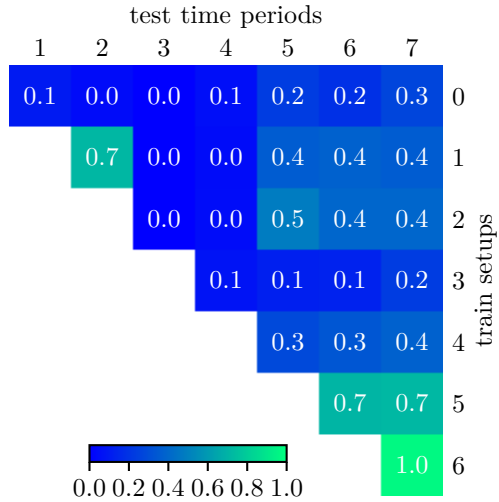

Figure 4. Performance of classifying selected iEEG clips into interictal/preictal classes as characterized by PR AUC values for dog B. For the meaning of time periods and train setups, refer to Tab. II and Fig. 1. Here a larger AUC value (green color) indicates a better performance.

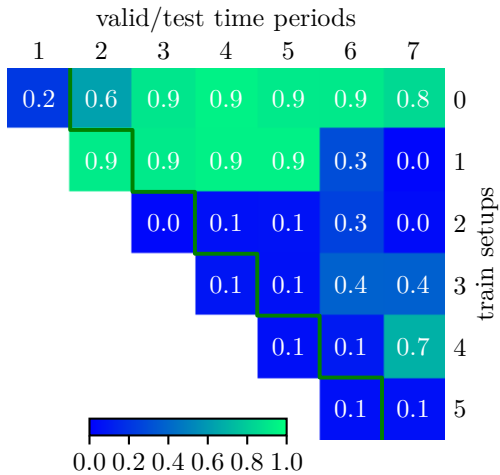

Figure 5. Performance of predicting seizure for continuous recording of dog A as characterized by TIW at 100%-sensitivity. The time period left adjacent to the green line was used to optimize the parameters of alarm setting. For the meaning of time periods and train setups, refer to Tab. II and Fig. 1. Here a smaller TIW value (blue color) indicates a better performance.

## IV. Conclusion

Using a machine learning algorithm by applying multilayer perceptrons and hand-crafted features, we have focused on how non-stationarity of iEEG time series affects the prediction performance and what clues can be found for the future design of the algorithm. Thereby we found that:

1) When varying the time period used for training the algorithms, one can observe significant changes in the prediction performance of most test recording periods, which points to the long-term evolution of cortical states on a time scale of weeks to months. Retraining of prediction algorithms is therefore necessary to better predict the impending seizures.
2) Seizures within their associated recording periods are intrinsically organized into groups, members of a group may even be far apart in time. The test performance can benefit significantly from train data from the same group. On the contrary, the inclusion of training data from other, different groups worsens the performance. These are clear indications that it is not the amount of train data improves the performance, but the composition of the train data.
3) The grouping of seizures within their associated recording time periods implies the existence of metastates of cortical dynamics. Thus, by applying a multi-state model, a coarse-grained representation of the temporal evolution of cortical dynamics can be obtained as the switching between such meta-states.
4) The retraining of prediction algorithms must be scheduled according to the change in cortical metastates. Furthermore, the composition of a train set should be adjusted as well.
5) There are evidences that the nature of seizure-free interictal data clips changes along with the transition between meta-states. The switch of cortical metastates and the right time for schedule a retraining may be probed via such changes in the interictal states.

### Acknowledgment

This work was supported by the European Regional Development Fund (ERDF), the Free State of Saxony (project number: 100320557), the Innovation Projects MedTech ALERT of Else Kröner Fresenius Center (EKFZ) for Digital Health of the TU Dresden and the University Hospital Carl Gustav Carus.

We thank Dhara Bhavsar for supporting the data download and the Center for Information Services and High Performance Computing (ZIH) at TU Dresden for generous allocation of computing time.

## References

[1] A. Schulze-Bonhage and A. Kühn, "Unpredictability of seizures and the burden of epilepsy," *Seizure prediction in epilepsy: from basic mechanisms to clinical applications*, pp. 1–10, 2008.

[2] L. Kuhlmann, K. Lehnertz, M. P. Richardson, B. Schelter, and H. P. Zaveri, "Seizure prediction - ready for a new era." *Nature reviews. Neurology*, vol. 14, pp. 618–630, Oct. 2018.

[3] U. R. Acharya, Y. Hagiwara, and H. Adeli, "Automated seizure prediction," *Epilepsy & Behavior*, pp. 251–261, nov.

[4] H. Daoud and M. A. Bayoumi, "Efficient Epileptic Seizure Prediction Based on Deep Learning," *IEEE Transactions on Biomedical Circuits and Systems*, vol. 13, no. 5, pp. 804–813, oct 2019.

[5] M. Hejazi and A. Motie Nasrabadi, "Prediction of epilepsy seizure from multi-channel electroencephalogram by effective connectivity analysis using Granger causality and directed transfer function methods," *Cognitive Neurodynamics*, pp. 1–13, may.

[6] H. Khan, L. Marcuse, M. Fields, K. Swann, and B. Yener, "Focal onset seizure prediction using convolutional networks," *IEEE Transactions on Biomedical Engineering*, may.

[7] K. M. Tsiouris, V. C. Pezoulas, M. Zervakis, S. Konitsiotis, D. D. Koutsouris, and D. I. Fotiadis, "A Long Short-Term Memory deep learning network for the prediction of epileptic seizures using EEG signals," *Computers in Biology and Medicine*, vol. 99, pp. 24–37, aug 2018.

[8] M. J. Cook, T. J. O'Brien, S. F. Berkovic, M. Murphy, A. Morokoff, G. Fabinyi, W. D'Souza, R. Yerra, J. Archer, L. Litewka, S. Hosking, P. Lightfoot, V. Ruedebusch, W. D. Sheffield, D. Snyder, K. Leyde, and D. Himes, "Prediction of seizure likelihood with a long-term, implanted seizure advisory system in patients with drug-resistant epilepsy: a first-in-man study," *The Lancet Neurology*, vol. 12, no. 6, pp. 563–571, jun 2013.

[9] B. H. Brinkmann, J. Wagenaar, D. Abbot, P. Adkins, S. C. Bosshard, M. Chen, Q. M. Tieng, J. He, F. J. Muñoz-Almaraz, P. Botella-Rocamora, J. Pardo, F. Zamora-Martinez, M. Hills, W. Wu, I. Korshunova, W. Cukierski, C. Vite, E. E. Patterson, B. Litt, and G. A. Worrell, "Crowdsourcing reproducible seizure forecasting in human and canine epilepsy," *Brain*, vol. 139, no. 6, pp. 1713–1722, jun 2016.

[10] I. Korshunova, P. J. Kindermans, J. Degrave, T. Verhoeven, B. H. Brinkmann, and J. Dambre, "Towards Improved Design and Evaluation of Epileptic Seizure Predictors," *IEEE Transactions on Biomedical Engineering*, 2018.

[11] L. Kuhlmann, P. Karoly, D. R. Freestone, B. H. Brinkmann, A. Temko, A. Barachant, F. Li, G. Titericz, B. W. Lang, D. Lavery, K. Roman, D. Broadhead, S. Dobson, G. Jones, Q. Tang, I. Ivanenko, O. Panichev, T. Proix, M. Náhlík, D. B. Grunberg, C. Reuben, G. Worrell, B. Litt, D. T. J. Liley, D. B. Grayden, and M. J. Cook, "Epilepsyecosystem.org: crowdsourcing reproducible seizure prediction with long-term human intracranial EEG," *Brain*, vol. 141, no. 9, pp. 2619–2630, aug 2018.

[12] C. Reuben, P. Karoly, D. R. Freestone, A. Temko, A. Barachant, F. Li, G. Titericz, B. W. Lang, D. Lavery, K. Roman, D. Broadhead, G. Jones, Q. Tang, I. Ivanenko, O. Panichev, T. Proix, M. Náhlík, D. B. Grunberg, D. B. Grayden, M. J. Cook, and L. Kuhlmann, "Ensembling crowdsourced seizure prediction algorithms using long-term human intracranial EEG," *Epilepsia*, p. epi.16418, dec 2019.

[13] K. A. Davis, B. K. Sturges, C. H. Vite, V. Ruedebusch, G. Worrell, A. B. Gardner, K. Leyde, W. D. Sheffield, and B. Litt, "A novel implanted device to wirelessly record and analyze continuous intracranial canine EEG," *Epilepsy Research*, vol. 96, no. 1-2, pp. 116–122, sep 2011.

[14] F. Mormann, R. G. Andrzejak, C. E. Elger, and K. Lehnertz, "Seizure prediction: the long and winding road," *Brain*, vol. 130, no. 2, pp. 314–333, 2007.

[15] F. Mormann and R. G. Andrzejak, "Seizure prediction: making mileage on the long and winding road," *Brain*, vol. 139, no. 6, pp. 1625–1627, may 2016.

[16] R. Tetzlaff and V. Senger, "The seizure prediction problem in epilepsy: Cellular nonlinear networks," *Circuits and Systems Magazine, IEEE*, vol. 12, no. 4, pp. 8–20, 2012.

[17] F. Mormann, K. Lehnertz, P. David, and C. E. Elger, "Mean phase coherence as a measure for phase synchronization and its application to the EEG of epilepsy patients," *Physica D: Nonlinear Phenomena*, vol. 144, no. 3, pp. 358–369, 2000.

[18] M. Eberlein, J. Müller, H. Yang, S. Walz, J. Schreiber, R. Tetzlaff, S. Creutz, O. Uckermann, and G. Leonhardt, "Evaluation of machine learning methods for seizure prediction in epilepsy," *Current Directions in Biomedical Engineering*, vol. 5, no. 1, pp. 109–112, 2019.

[19] J. Müller, H. Yang, M. Eberlein, G. Leonhardt, O. Uckermann, L. Kuhlmann, and R. Tetzlaff, "Coherent false seizure prediction in epilepsy: Coincidence or providence?" *Submitted to Clinical Neurophysiology*, 2021.

[20] M. Eberlein, R. Hildebrand, R. Tetzlaff, N. Hoffmann, L. Kuhlmann, B. Brinkmann, and J. Müller, "Convolutional Neural Networks for Epileptic Seizure Prediction," in *Proc. IEEE Int. Conf. Bioinformatics and Biomedicine (BIBM)*, Dec. 2018, pp. 2577–2582.

[21] "Genetic algorithms for real parameter optimization," ser. Foundations of Genetic Algorithms, G. J. RAWLINS, Ed. Elsevier, 1991, vol. 1, pp. 205–218.